\documentclass[twocolumn]{aastex62}
\usepackage[section]{placeins}
\usepackage{graphicx}
\usepackage{hyperref}
\usepackage[caption=false]{subfig}

\usepackage{float}

\shorttitle{Dynamical Modeling of CX137}
\shortauthors{Gomez et al.}

\begin{document}
 
\title{Dynamical Modeling of CXOGBS J175553.2-281633: A 10 Hour Long Orbital Period Cataclysmic Variable}

\correspondingauthor{Sebastian Gomez}
\email{sgomez@cfa.harvard.edu}

\author[0000-0001-6395-6702]{Sebastian Gomez}
\affil{Center for Astrophysics \textbar{} Harvard \& Smithsonian, 60 Garden Street, Cambridge, MA, 02138, USA}

\author[0000-0002-5297-2683]{Manuel A.~P.~Torres}
\affil{Instituto de Astrofísica de Canarias, E-38205 La Laguna, S/C de Tenerife, Spain}
\affil{Departamento de Astrof\'\i{}sica, Universidad de La Laguna, E-38206 La Laguna, Tenerife, Spain}

\author[0000-0001-5679-0695]{Peter G.~Jonker}
\affil{SRON, Netherlands Institute for Space Research, Sorbonnelaan 2, 3584 CA Utrecht, The Netherlands}
\affil{Department of Astrophysics/IMAPP, Radboud University,
P.O.~Box 9010, 6500 GL, Nijmegen, The Netherlands}

\author{Zuzanna Kostrzewa-Rutkowska}
\affil{SRON, Netherlands Institute for Space Research, Sorbonnelaan 2, 3584 CA Utrecht, The Netherlands}
\affil{Department of Astrophysics/IMAPP, Radboud University,
P.O.~Box 9010, 6500 GL, Nijmegen, The Netherlands}

\author[0000-0002-1206-1930]{Theo F.~J.~van Grunsven}
\affil{SRON, Netherlands Institute for Space Research, Sorbonnelaan 2, 3584 CA Utrecht, The Netherlands}
\affil{Department of Astrophysics/IMAPP, Radboud University,
P.O.~Box 9010, 6500 GL, Nijmegen, The Netherlands}

\author[0000-0001-5207-5619]{Andrzej Udalski}
\affil{Astronomical Observatory, University of Warsaw, Al. Ujazdowskie 4, 00-478 Warszawa, Poland}

\author{Robert I. Hynes}
\affil{Department of Physics and Astronomy, Louisiana State University, Baton Rouge, Louisiana 70803, USA}

\author[0000-0003-3944-6109]{Craig O. Heinke}
\affil{Dept. of Physics, University of Alberta, CCIS 4-183, Edmonton, AB, T6G 2E1, Canada}

\author[0000-0003-0976-4755]{Thomas J. Maccarone}
\affil{Department of Physics \& Astronomy, Texas Tech University, Box 41051, Lubbock TX 79409, USA}

\author[0000-0002-1206-1930]{Ricardo Salinas}
\affil{Gemini Observatory/NSF's National Optical-Infrared Astronomy Research Laboratory, Casilla 603, La Serena, Chile}

\author[0000-0002-1468-9668]{Jay Strader}
\affil{Center for Data Intensive and Time Domain Astronomy, Department of Physics and Astronomy, Michigan State University, East Lansing, MI 48824, USA}

\begin{abstract}

We present modeling of the long-term optical light curve and radial velocity curve of the binary stellar system CXOGBS J175553.2-281633, first detected in X-rays in the \textit{Chandra} Galactic Bulge Survey. We analyzed \mbox{7 years} of optical I-band photometry from OGLE and found long-term variations from year to year. These long-term variations can most likely be explained with by either variations in the luminosity of the accretion disk or a spotted secondary star. The phased light curve has a sinusoidal shape, which we interpret as being due to ellipsoidal modulations. We improve the orbital period to be \mbox{$P = 10.34488 \pm 0.00006$ h} with a time of inferior conjunction of the secondary star $T_0 = {\rm HJD\ } 2455260.8204 \pm 0.0008$. Moreover, we collected 37 spectra over 6 non-consecutive nights. The spectra show evidence for an evolved K7 secondary donor star, from which we obtain a semi-amplitude for the radial velocity curve of \mbox{$K_2 = 161 \pm 6 $ km s$^{-1}$}. Using the light curve synthesis code {\tt XRbinary}, we derive the most likely orbital inclination for the binary of \mbox{$i = 63.0\pm0.7$ deg}, a primary mass of \mbox{$M_1 = 0.83 \pm 0.06$ M$_\odot$}, consistent with a white dwarf accretor, and a secondary donor mass of \mbox{$M_2 = 0.65 \pm 0.07$ M$_\odot$}, consistent with the spectral classification. Therefore, we identify the source as a long orbital period cataclysmic variable star.

\end{abstract}


\section{Introduction} \label{sec:intro}

Cataclysmic variables (CVs) are binary star systems composed of a white dwarf primary accreting matter from a main sequence or evolved secondary star \citep{Patterson84,Warner95,Kalomeni16}. Low mass X-ray binaries (LMXBs) are analogous systems where the primary is either a black hole or a neutron star, instead of a white dwarf \citep{Remillard06}. There are only about 20 dynamically confirmed black hole X-ray binaries known in the Milky Way (e.g., \citealt{Casares14}). Finding and modelling CVs and LMXBs allows us to better understand the formation of compact objects and test binary evolution models \citep{Jonker04,Repetto15}.

The \textit{Chandra} Galactic Bulge Survey (GBS) is a survey tasked with finding more quiescent LMXBs. Towards this goal the GBS covered a total of 12 deg$^2$ near the Bulge of our galaxy and found 1640 X-ray sources \citep{Jonker11,Jonker14}. Subsequent studies have identified counterparts to these sources in multiple wavelengths; from radio \citep{Maccarone12} and near infrared \citep{Greiss14} to optical \citep{Hynes12, Udalski12, Britt14, Wevers16, Wevers17}. Some of these counterparts have been deemed likely accreting binaries, motivating further photometric and spectroscopic follow up \citep{Ratti13, Wevers16_CVn, Johnson17}.

In this work we focus on one of these objects, CXOGBS J175553.2-281633 (hereafter CX137). The optical counterpart to CX137 was first identified by \cite{Udalski12} and classified as an eclipsing binary with a spotted donor star and an orbital period of $10.345$ hr. Subsequent spectroscopic and photometric follow up by \cite{Torres14} revealed broad H$\alpha$ emission and an orbital period consistent with that of \cite{Udalski12}. Based on the properties of the H$\alpha$ emission line and an X-ray luminosity of \mbox{$L_x > 5.8\times10^{30}$ erg s$^{-1}$}, \cite{Torres14} classified the source as a potential low-accretion rate CV or quiescent LMXB with a G/K-type secondary, supporting the ellipsoidal light curve interpretation.

In this work we build on the analysis from \cite{Torres14} by including two extra years of I-band photometry, where the sinusoidal shape of the light curve can be explained by ellipsoidal modulations. Additionally, we see long-term variations in the shape of the light curve, these are consistent with either changes in the luminosity of an accretion disk or a spotted secondary star. In this work we aim to settle the true nature of the object by performing a dynamical study and find that CX137 is a CV with a K7 secondary star and an orbital period of \mbox{$P = 10.34488 \pm 0.00006$ h}, in agreement with previous studies. The source shows no outbursts in our seven years of optical photometry.

This paper is organized as follows: in \S\ref{sec:data} we describe the OGLE photometry and the optical spectroscopy obtained for this study. In \S\ref{sec:data_analysis} we provide an analysis of the data; where we determine the orbital period, generate a radial velocity curve for the secondary star, and describe the spectral features. In \S\ref{sec:modeling} we present our light curve models, fitting routines and resulting output parameters. We finally outline our discussion in \S\ref{sec:params} and conclusion in \S\ref{sec:conclusions}. All quoted errors in this paper represent $1\sigma$ uncertainty, unless otherwise stated.

\section{Observations} \label{sec:data}

\subsection{Gaia} \label{subsec:gaia}

\textit{Gaia} provides precise coordinates for the optical component of CX137 at \mbox{R.A.=${\rm 17^h55^m53^s.26}$}, \mbox{decl.=$-28^\circ16'33''.84$} (ICRS), in addition to proper motion components of \mbox{$\mu_{\rm R.A.} = 1.139\pm0.108$ mas yr$^{-1}$}, and \mbox{$\mu_{\rm decl.}=-6.977\pm0.087$ mas yr$^{-1}$}. The parallax of the source was measured by \textit{Gaia} DR2 to be \mbox{$\pi = 1.116 \pm 0.069$ mas}, which corresponds to a distance of \mbox{$d = 879^{+59}_{-52}$ pc} \citep{Bailer18, Gaia18}.

\begin{figure}
	\centering
	\includegraphics[width=\columnwidth]{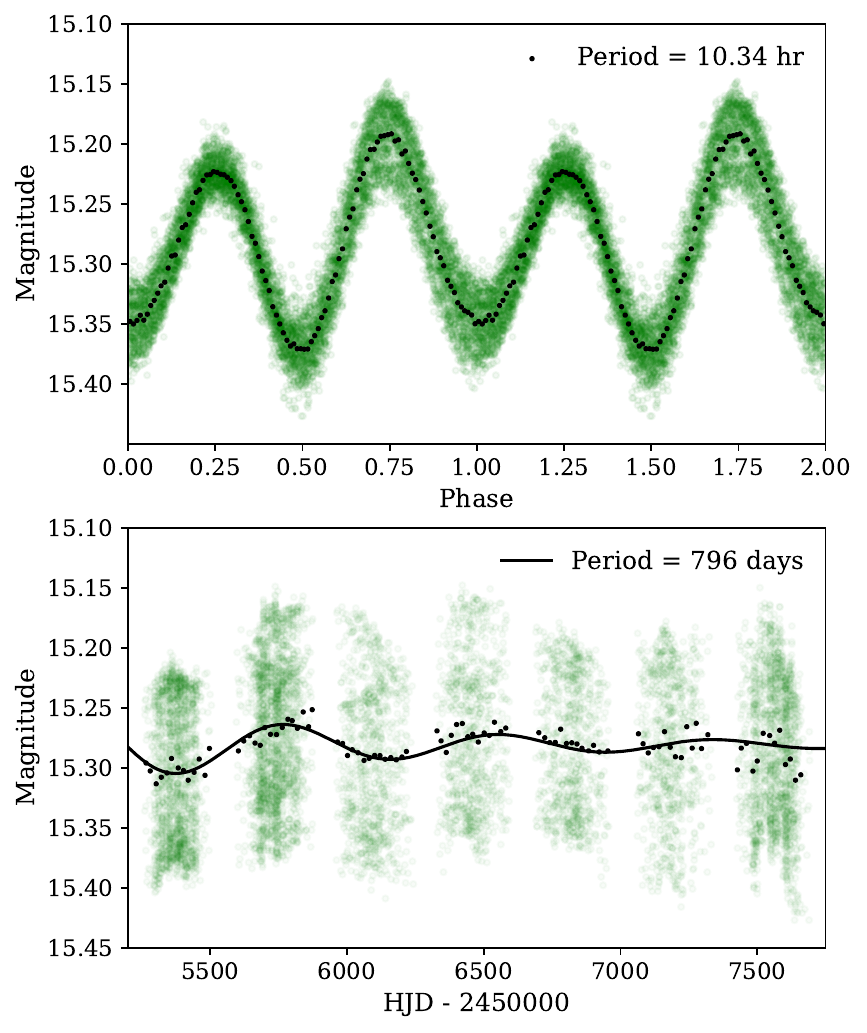}
	\caption{\textit{Top:} Optical photometry phased at the best orbital period of $P = 10.34488$ h. \textit{Bottom:} Full OGLE Light curve where long-term periodic variations in luminosity are seen. The green dots are all the data, while the black dots are binned in phase bins of 0.01 for the phased light curve and bins of 20 days for the full light curve. We show a tentative period of 796 days as a damped sine curve fit to the binned data. The error bars are approximately equal to the size of the data points and are not plotted for clarity. \label{fig:phased}}
\end{figure}

\subsection{OGLE Photometry} \label{subsec:photometry}

The optical counterpart of CX137 was observed during the fourth phase of the Optical Gravitational Lensing Experiment (OGLE) project with the 1.3m Warsaw telescope at Las Campanas Observatory \citep{Udalski15}. OGLE provided us with 7 years of \mbox{I-band} photometry, from 2010 to 2016. The typical cadence of these observations ranges from 20 minutes to nominally once a night with exposure times of 100s. There is a three month period in each year when the source is not visible. The photometry was obtained using the difference image analysis method outlined in \cite{Wozniak00}. The individual photometry has typical errors of $<0.01$ mag, see \mbox{Table~\ref{tab:photometry}} for a log of observations.

\begin{figure}
	\centering
	\includegraphics[width=\columnwidth]{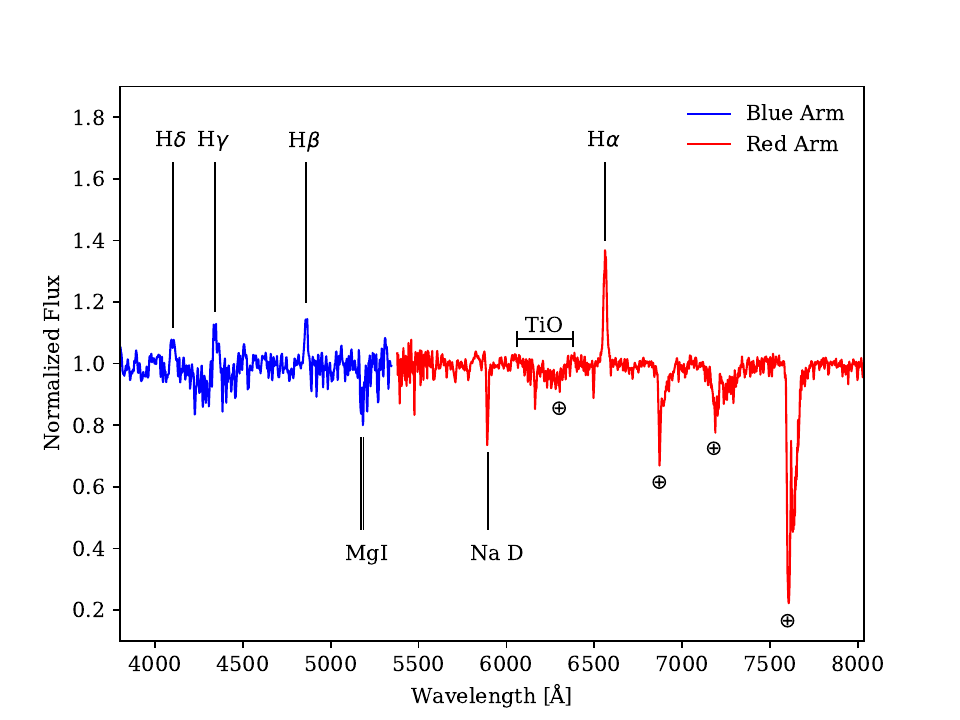}
	\caption{Average continuum-normalized blue and red arm ISIS spectrum for CX137 in the rest frame of the secondary. The spectra show Balmer lines in emission, associated with the accretion disk. Strong stellar features are indicated. The interstellar Na\,D is also marked. $\oplus$ denotes prominent telluric features. \label{fig:spectra}}
\end{figure}

\subsection{Optical Spectroscopy} \label{subsec:spectra}

We observed CX137 with the Intermediate dispersion Spectrograph and Imaging System (ISIS; \citealt{Jorden90}) on the 4.2 m William Herschel Telescope (WHT) at the Roque de los Muchachos Observatory on La Palma, Spain, during 5 different observing runs between June and August 2017. The ISIS spectrograph has a dichroic that splits the spectra into a red and blue arm, allowing for a wide spectral range to be observed simultaneously. For the blue arm we used gratings R158, R300, and R600; and for the red arm we used gratings of R158 and R600. We also obtained one high resolution spectrum with the Inamori-Magellan Areal Camera and Spectrograph (IMACS; \citealt{dressler11}) on the Magellan Baade 6.5 m Telescope at Las Campanas observatory with the R1200 grating. We provide a log of spectroscopic observations and specifications of each grating in Table~\ref{tab:spectroscopy}. The spectral resolutions provided in the table were approximated by measuring the width of spectroscopic lines in an arc lamp spectrum taken with each grating.

We reduced the spectra using standard IRAF\footnote{\label{IRAF}IRAF is written and supported by the National Optical Astronomy Observatories, operated by the Association of Universities for Research in Astronomy, Inc. under cooperative agreement with the National Science Foundation.} routines. The data  were bias-subtracted and flat-fielded, sky emission subtracted, the spectra were optimally extracted and wavelength calibrated using an arc lamp taken after each spectrum. We determine the zero-point of the wavelength calibration of our spectra by measuring the positions of bright sky lines in each spectrum, and apply the corresponding shift to each individual spectrum such that the wavelength of the sky lines match between all the spectra. For the ISIS spectroscopy, we analyzed the data taken in both the red and blue arms using the same procedure, but treat them as individual spectra.

\begin{figure}
	\centering
	\includegraphics[width=\columnwidth]{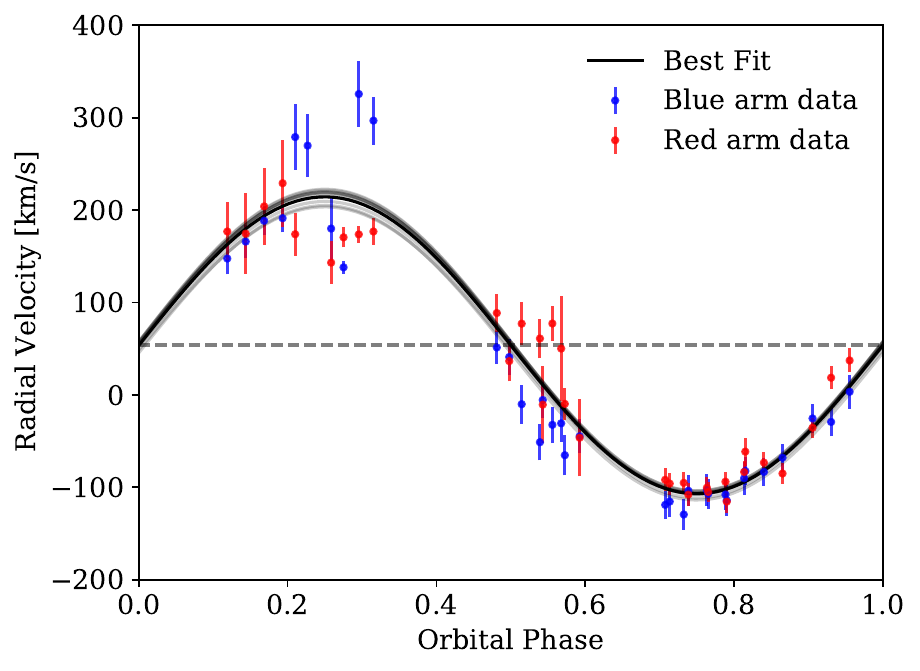}
	\caption{Heliocentric radial velocity curve measured from both the blue and red arm of ISIS spectra. The best fit sine curve is shown in black. The dashed horizontal line marks the 0 point of the sinusoid. The best fit values to the systemic radial velocity and semi-amplitude of the radial velocity are \mbox{$\gamma = 54 \pm 4$ km s$^{-1}$} and K$_2 = 161 \pm 6$ km s$^{-1}$, respectively. \label{fig:rvfit}}
\end{figure}

\subsection{Spectral Templates} \label{subsec:templates}

Throughout this work we make use of spectral templates from the X-shooter library \citep{Chen11}. We selected spectra from 71 M stars, 33 K stars and 23 G stars of varying luminosity classes and evolutionary stages. All templates were taken with a $0.''7$ slit with the VIS arm of X-shooter and a nominal resolution of \mbox{R $\sim 10,000$}, equivalent to $\sim30$ km s$^{-1}$ at a wavelength of 8600 \AA.

All the spectra of CX137 and templates were subsequently processed using {\tt Molly}\footnote{{\tt Molly} is a code developed and maintained by T. Marsh and it is available at \url{http://deneb.astro.warwick.ac.uk/phsaap/software/molly/html/INDEX.html}\label{foot:molly}}. First, we apply a heliocentric velocity correction to all spectra using the \textit{hfix} task. We then use \textit{vbin} to bin all the data to a uniform velocity scale so the dispersion of the templates matches that of the CX137 spectra. We then normalize each spectrum by dividing it by a fit to the star's continuum. To estimate the continuum we fit a 2nd-order polynomial to each spectrum, masking out regions with strong emission lines or telluric bands.

\begin{deluxetable}{cc}
	\tablecaption{OGLE I-band photometry. All exposure times were 100s. \label{tab:photometry}}
	\tablehead{\colhead{\hspace{0.5cm}UT Date Range\hspace{0.5cm}} & \colhead{\hspace{0.5cm}\ Exposures}\hspace{0.5cm}} 
	\startdata
	Mar 5 - Nov 4  2010 & 1685 \\
	Feb 3 - Nov 9  2011 & 2042 \\
	Feb 3 - Nov 11 2012 & 936  \\
	Feb 3 - Oct 30 2013 & 868  \\
	Feb 1 - Oct 26 2014 & 848  \\
	Feb 7 - Nov 7  2015 & 804  \\
	Feb 6 - Oct 30 2016 & 1641 \\
	\enddata
\end{deluxetable}

\section{Data Analysis} \label{sec:data_analysis}

\begin{deluxetable*}{ccccccccc}
    \tablecaption{Optical Spectroscopy of CX137 \label{tab:spectroscopy}}
    \tablewidth{0pt}
    \tablehead{
        \colhead{UT Date}   & \colhead{Exposure} & \colhead{Telescope +} & \colhead{Grating} & \colhead{Dispersion} & \multicolumn{2}{c}{Resolution} & \colhead{Slit width} & \colhead{Wavelength range} \\
                            & \colhead{(s)}      & \colhead{Instrument}  & (lines/mm)        & (\AA\ pixel$^{-1}$)  & (\AA)                & (km s$^{-1}$)        & (arcsec)             & (\AA)}
    \startdata
    2017 Jun 24             & $6  \times 600$    & WHT + ISIS-red        & R158              & 1.81                 & 7.70                 & 307                  & 1.0"                 & 5500 - 8100                \\
    2017 Jun 24             & $6  \times 600$    & WHT + ISIS-blue       & R158              & 1.62                 & 7.81                 & 520                  & 1.0"                 & 3500 - 5400                \\
    2017 Jul 11             & $6  \times 600$    & WHT + ISIS-red        & R158              & 1.81                 & 7.70                 & 307                  & 1.0"                 & 5500 - 8100                \\
    2017 Jul 11             & $6  \times 600$    & WHT + ISIS-blue       & R300              & 0.86                 & 4.10                 & 273                  & 1.0"                 & 3500 - 5400                \\
    2017 Jul 21             & $15 \times 900$    & WHT + ISIS-red        & R600              & 0.49                 & 1.81                 & 72                   & 1.0"                 & 5500 - 8800                \\
    2017 Jul 21             & $15 \times 900$    & WHT + ISIS-blue       & R600              & 0.45                 & 2.02                 & 134                  & 1.0"                 & 3500 - 5400                \\
    2017 Aug 27             & $3  \times 900$    & WHT + ISIS-red        & R600              & 0.49                 & 1.81                 & 72                   & 1.0"                 & 5500 - 7150                \\
    2017 Aug 27             & $3  \times 900$    & WHT + ISIS-blue       & R600              & 0.45                 & 2.02                 & 134                  & 1.0"                 & 3910 - 5400                \\
    2017 Aug 29             & $4  \times 900$    & WHT + ISIS-red        & R600              & 0.49                 & 1.81                 & 72                   & 1.0"                 & 5500 - 7150                \\
    2017 Aug 29             & $4  \times 900$    & WHT + ISIS-blue       & R600              & 0.45                 & 2.02                 & 134                  & 1.0"                 & 3910 - 5400                \\
    2017 Oct 8              & $3  \times 900$    & Magellan + IMACS      & R1200             & 0.376                & 1.54                 & 54                   & 0.9"                 & 8500 - 8900                \\
    \enddata
    \tablecomments{The spectral resolution is measured at 4500 \AA\ for the ISIS-blue arm, 7500 \AA\ for the ISIS-red arm, and 8600 \AA\ for IMACS. The wavelength range represents only the high quality portion of the spectra used for our analysis.}
\end{deluxetable*}

\subsection{Photometric Periodicities} \label{sec:period}

We use all 7 years of OGLE I-band photometry to determine the orbital period of the binary. For this we employ the {\tt gatspy} python package \citep{VanderPlas15}, which provides an implementation of the Lomb-Scargle periodogram to find periodicities in the photometric data. The strongest peak of the periodogram is at a period of $P = 5.17244$ h. When the data are phase-folded at this period we see large scatter in the light curve, which is due to the fact that the maxima expected from ellipsoidal modulations at phase 0.25 and 0.75 have different strengths (see Figure~\ref{fig:phased}). Figure~\ref{fig:phased} shows the light curve phase-folded at a period of twice that of the corresponding strongest peak, $P = 10.34488 \pm 0.00006$ hr, consistent with the period found in \cite{Udalski12} and \cite{Torres14}. Motivated by the fact that spin periods in the range of $0.1-10$\% of the orbital period have previously been observed in magnetic CVs \citep{Norton04}, we search for periodicities in the 100--20,000 s range with null results. We detect no measurable change in the orbital period over our 7 year baseline. On the other hand, we detect a possible long-term trend at a period of $\sim 796$ days. Since the full span of the light curve is only three times this period, more data are required to confirm if this is a real periodicity or just a temporary artifact. The data phase-folded at this period is shown in the bottom panel of Figure~\ref{fig:phased}.

\subsection{Spectral Type of the Secondary} \label{sec:spectra}

Figure~\ref{fig:spectra} shows the blue and red normalized ISIS spectra of CX137 averaged in the rest frame of the secondary star. The spectra are mostly dominated by absorption lines from the secondary, with additional Balmer emission lines from an accretion disk. We detect the Mg triplet absorption lines from the secondary, and interstellar Na D lines. We see a weak contribution from TiO bands of the secondary in the $\sim 6100 - 6300$ \AA\ range, and no evidence for \ion{He}{1} emission lines, which are common in CVs (e.g, \citealt{Ratti13, Rodriguez09}). This might be due to the lines being veiled by a large flux contribution from the secondary. We can set an upper limit to the absolute equivalent width of \ion{He}{1} $\lambda7065$ to be $< 1.6$ \AA, and $<1.2$ \AA\ for \ion{He}{2} $\lambda4686$.

To estimate the temperature of the secondary star we first average all the CX137 ISIS data taken with the R600 grating to use as a high S/N reference. We compare this spectrum to that of the X-shooter templates described in Section~\ref{subsec:templates}. First, we corrected each template spectrum for the systemic velocity of each star and broaden it by convolving it with a Gaussian function to match the spectral resolution of the ISIS data. We subtract each template to the normalized CX137 spectrum in the $ 5580 - 6150$ \AA\ wavelength range (masking out telluric lines and emission lines not associated with the secondary), and search for the template star that produces the lowest residuals, allowing for a varying multiplicative $f$ factor, which represents the fractional contribution of the template star from the total flux. We find that the spectrum of CX137 best matches that of HD79349, a K7IV star with a temperature of $3850 \pm 30$K, and a systemic velocity of \mbox{$47.12 \pm 0.15$ km s$^{-1}$} \citep{Arentsen19, Gaia18}. We find a best fit for the average optimum factor of $f = 0.52 \pm 0.06$.

\subsection{Radial Velocities} \label{subsec:RV}

To measure the radial velocity of the secondary in each spectrum we use the \textit{xcor} task in {\tt Molly} to cross-correlate the CX137 spectra with the spectrum of the K7IV star HD79349, the template star that best matches the spectra of CX137 (described in Section~\ref{sec:spectra}). The actual choice of template star does not have a noticeable effect in the measured radial velocities. We correct the template star's spectrum for its systemic velocity and broaden it by convolving it with a Gaussian function to match the spectral resolution of the CX137 spectra. We consider the wavelength range listed in Table~\ref{tab:spectroscopy} for each CX137 spectrum, masking out telluric features and emission lines not associated with the secondary before cross-correlating them. We calculate the radial velocities from both the red and blue arm data of the ISIS spectrograph independently. The resulting radial velocity curve is shown in Figure~\ref{fig:rvfit}, with the individual measurements provided in Table~A.\ref{tab:radial_velocities}. We note that the radial velocities measured near phase $0.25$ have a large scatter due to noisy spectra taken in poor weather conditions.

We model the radial velocity curve with a sine function of the form:

\begin{equation}\label{eq:rv}
v \left(t\right) = \gamma + K_2\sin \left[2\pi \left( \frac{t}{P} + \phi \right) \right],
\end{equation}
\noindent
where we fix the orbital period to be $P = 10.34488$ h, as determined in section~\ref{sec:period}. We fit for the radial velocity semi-amplitude $K_2$, a systemic velocity $\gamma$, and a phase offset $\phi$, where $\phi = 0$ corresponds to the photometric phase 0, or inferior conjunction of the secondary star. We find a best fit model with \mbox{K$_2 = 161 \pm 6$ km s$^{-1}$}, \mbox{$\gamma = 54 \pm 4$ km s$^{-1}$}, and $\phi = 0.00 \pm 0.02$ with a corresponding $\chi^2 = 141$ and 64 degrees of freedom. The quoted uncertainties are for a model where we scale the error bars of the individual radial velocity measurements to correspond to a reduced $\chi^2 = 1$ (e.g., \citealt{Marsh94}). The error-bars shown in Figure~\ref{fig:rvfit} are the true measured error-bars, not scaled.

\subsection{Rotational Broadening of the Secondary Star} \label{subsec:RB}

To estimate the rotational broadening of the secondary we compare the set of spectral templates described in Section~\ref{subsec:templates} to the high resolution IMACS spectrum of CX137 taken near photometric orbital phase 0. We normalize the IMACS and X-shooter spectra by dividing them by a 2nd degree polynomial fit to their respective continuum (masking out absorption features) in the $8500-8900$ \AA\ range. We scale down the resolution of the X-shooter templates to match that of the IMACS spectrum by convolving them with a Gaussian function. We then broaden the templates by a range of velocities from $20-200$ km s$^{-1}$ in steps of 1 km s$^{-1}$ using the \textit{rbroad} task in {\tt Molly}. This task takes the input spectrum and broadens it through convolution with the rotational profile of \cite{Gray92}, where we adopt a limb darkening coefficient of 0.75. Finally, we subtract the broadened templates from the CX137 spectrum, following the same procedure as described in \S\ref{sec:spectra}. Through $\chi^2$ minimization we find a best fit of \mbox{$v \sin(i) = 101 \pm 3$ km s$^{-1}$} to the rotational velocity of the secondary star in CX137. We find the best match to be comparably good for a K4III, K3.5III, K2III, and K7IV template star. G and M stars produce statistically worst fits. The individual $v\sin(i)$ measurements are shown in Table~\ref{tab:templates}.

From the \mbox{$v \sin(i) = 101 \pm 3$ km s$^{-1}$} estimate and velocity semi-amplitude \mbox{$K_2 = 161 \pm 6$ km s$^{-1}$} calculated in section~\ref{subsec:RV} we obtain a mass ratio of \mbox{$q = M_2 / M_1 = 0.79 \pm 0.06$} using equation~\ref{eq:vsiniq}, which holds for a Roche Lobe filling secondary that co-rotates with the binary orbit \citep{Wade88}.

\begin{equation}\label{eq:vsiniq}
\frac{v \sin(i)}{K_2} = 0.462 [(1 + q)^2q]^{1/3}
\end{equation}

\begin{figure*}
	\centering
	\includegraphics[width=0.85\textwidth]{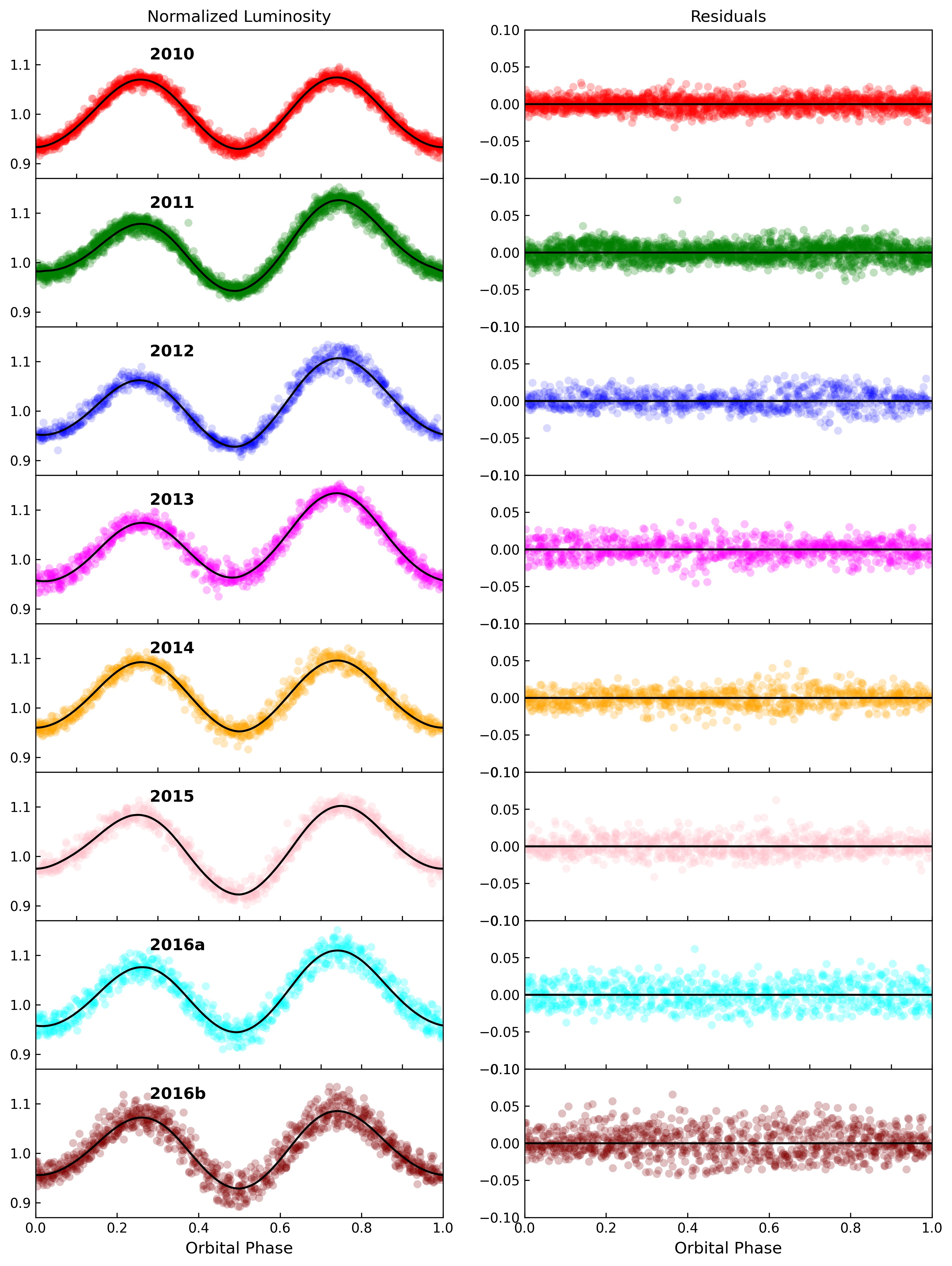}
	\caption{\textit{Left} : Optical light curves for all eight epochs of observations of CX137 phased at the photometric ephemeris. We include the best fit model described in Section~\ref{sec:spots} in black, where the luminosity of each epoch is divided by its average luminosity. Each panel shows a different epoch in order of time, error bars are not plotted since they are smaller than the data marker size. \textit{Right} : Fractional residuals of the best-fit model to the light curve. \label{fig:model}}
\end{figure*}

\section{Light Curve Modeling} \label{sec:modeling}

\begin{deluxetable}{cccc}
	\tablecaption{Rotational broadening of CX137 for different templates \label{tab:templates}}
	\tablewidth{0pt}
	\tablehead{
		\colhead{Star} & \colhead{Spectral Type} & \colhead{v$\sin(i)$}   & \colhead{$f$}  \\
		 		       &                         &  \colhead{km s$^{-1}$} &                           }
	\startdata
    HD37763      & K2III    & $99  \pm 3$ & $0.48 \pm 0.06$ \\
    HD79349      & K7IV     & $100 \pm 3$ & $0.41 \pm 0.03$ \\
    BS4432       & K3.5III  & $100 \pm 3$ & $0.52 \pm 0.04$ \\
    HD74088      & K4III    & $104 \pm 3$ & $0.50 \pm 0.04$ \\
	\enddata
    \tablecomments{$f$ is the corresponding optimum factor measured in the $8500-8900$ \AA\ range.}
\end{deluxetable}

We proceed to model the optical light curve of CX137 using {\tt XRbinary}\footnote{A full description of {\tt XRbinary} can be found at \url{http://www.as.utexas.edu/~elr/Robinson/XRbinary.pdf}}, a light curve synthesis code developed by E.L. Robinson. This code assumes a binary system composed of a compact primary and a co-rotating secondary star that fills its Roche Lobe and is transferring mass via an accretion disk. The code models the tidal distortion of the secondary (responsible for the ellipsoidal modulations), and accounts for irradiation of the surface of the secondary from the bright accretion disk. The accretion disk is assumed to be optically thick and to emit as a multi-temperature blackbody. The disk's temperature profile as a function of radius is given by an equation of the form $T^4 \propto R^{-3} \left(1 - (R_{\rm in} / R)^{0.5} \right)$, where $R_{\rm in}$ is the inner disk radius. In order to account for the observed symmetries in the light curve, we model the photometry of CX137 with three different models: (i) a model with a Roche Lobe filling secondary and an accretion disk that is allowed to vary in luminosity and eccentricity, (ii) a similar model, but with a circular accretion disk where the temperature of the edge of the disk can have a hot and a cool side, and (iii) a model with a circular accretion disk and an edge of uniform temperature, but with two spots on the surface of the secondary. For all models we fit the light curve using the {\tt emcee} MCMC sampler \citep{Foreman13}.

The relevant parameters of the model are: the inclination of the system $i$; an orbital phase shift of the photometric $T_0$ with respect to the spectroscopic $T_0$, $\phi$; the temperature of the secondary star $T_2$; the temperature of the edge of the accretion disk $T_E$; the mass ratio \mbox{$q = M_2 / M_1$}; the argument of periastron of the disk $\omega_D$; the outer disk radius $R_D$; the disk luminosity $L_D$; the height of the accretion disk $H_D$; the eccentricity of the accretion disk $e_D$; the temperature ratio between the hot and cool sides of the disk edge $T_h$; the width of the hot side of the disk edge $W_h$; the location of the center of the hot edge of the disk $\theta_h$; the polar coordinates of the first and second spot on the surface of the secondary $\phi_{S1}$, $\theta_{S1}$, $\phi_{S2}$, and $\theta_{S2}$, respectively; the temperature ratio between the spots' temperature and the secondary temperature $T_{S1}$, and $T_{S2}$ respectively, and the size of the spots $R_{S1}$, and $R_{S2}$. Only the relevant parameters are included in each of the three versions of the models described in the following section.

In all models we fix the semi-amplitude of the radial velocity of the secondary to \mbox{$K_2 = 161$ km s$^{-1}$} (derived in \S\ref{subsec:RV}). We use wide uniform priors for $\phi$, $T_2$, $T_E$, $\omega_D$, $R_D$, $e_D$, $W_h$, and all the parameters pertaining to the spots. For $L_D$ we use a prior that is flat in log space to allow for even sampling of the parameter space across orders of magnitude. For $i$ we use a prior that is flat in $\cos(i)$. We implement a Gaussian prior on the mass ratio centered at \mbox{$q = 0.79 \pm 0.06$} (derived in \S\ref{subsec:RB}). We restrict the accretion disk to be larger than the circularization radius $R_c = (1 + q) (0.5 - 0.227 \log(q))^4$ \citep{Frank02}. Finally, apply a flat prior on the temperature of the secondary of \mbox{$T_2 = [3500, 4100]$}, based on the temperature of the template star that best matches the spectra of CX137 (derived in Section~\ref{sec:spectra}). The {\tt XRbinary} code interpolates the temperature from a table of Kurucz models, therefore the measurement of the temperature of the secondary is not very precise ($\pm125$ K), we report only the statistical model uncertainties in Table~\ref{tab:models}.

In order to account for the year-to-year variations in the light curve we separate the photometry into eight epochs, nominally one for every year of data. Dividing the photometry into eight epochs allows us to roughly track the evolution of the system, assuming the parameters of the system are approximately constant in the $\sim 8$ months of data each epoch spans (see Figure~\ref{fig:phased}). We see the shape of the light curve does remain fairly constant within each epoch, except for the 2016 epoch, which we therefore split into two epochs of equal time span named 2016a and 2016b, each of which do have a stable light curve shape. Subdividing the epochs further proved to be too computationally expensive.

Given that we know the orbital period of the binary is \mbox{$P = 10.34488$ h} we can calculate the mass function according to the equation:

\begin{equation}\label{eq:massfunction}
\frac{M_2^3 \sin(i)^3}{(M_1 + M_2)^2} = \frac{P K_2^3}{2 \pi G},
\end{equation}

\noindent
where we are able to determine the primary and secondary mass of the system by reparametrizing $q$, $K_2$, and $i$ using equation~\ref{eq:vsiniq}. 

\subsection{Model 1 : Variable Disk}\label{sec:eccentric}

For the first model we allow the accretion disk to vary in luminosity and eccentricity, but do not include any spots on the disk or the secondary. For all epochs we keep $i$, $\phi$, $T_2$, $K_2$, $T_E$, and $q$ constant but allow the parameters that define the accretion disk $\omega_D$, $R_D$, $L_D$, and $e_D$, $H_D$ to vary from epoch to epoch. The temperature of the edge of the disk $T_E$ could conceivably change from epoch to epoch, but since this parameter has little to no effect on the output light curve we constrain it to always be the same for computational purposes.

First, we fit each epoch of photometry independently, we then use the posterior distribution of those MCMC chains as starting positions when fitting all eight epochs simultaneously. We run the MCMC sampler for 1600 steps with 400 walkers and discard the first 50\% as burn-in. We test for convergence by using the Gelman-Rubin statistic and see that the potential scale reduction factor is $\hat{R} < 1.3$ \citep{Gelman92}. The most likely values are shown in Table \ref{tab:parameters}. We see that the posterior distribution of all the relevant parameters is mostly Gaussian. 

For this model we interpret the changes in the light curve as being due to an accretion disk of varying shape and luminosity. We see the light curves are well modeled by a disk that gets smaller and more eccentric from 2010 to 2013, and then recedes back to its original luminosity 3 years later and circularizes into a less eccentric disk. The best fit parameters for each epoch are shown in Table~\ref{tab:parameters}. We find a best fit for the secondary temperature of $4055\pm25$ K and a secondary mass of $M_2 = 0.62\pm0.04$ M$_\odot$, both consistent with a K7 star \citep{Cox00} and in agreement with the spectral classification performed in Section~\ref{sec:spectra}. The large Roche Lobe radius of the secondary $R_2 = 0.92 \pm 0.09$ R$_\odot$ implies it must be evolved in order to fill its Roche Lobe (discussed further in \S\ref{sec:params}).

We note that some of the best-fit eccentricity measurements are as high as $e = 0.58$, which is not expected for a low accretion-rate CV with a accretion disk of radius $\sim R_c$, and for the high mass ratio found for CX137 of $q>0.7$ (e.g, \citealt{Warner95}). For this reason we proceed to model the light curve with a disk that is forced to be circular, but with an edge that has two zones of independent temperature.

\subsection{Model 2 : Asymmetrical Disk Edge Brightness}\label{sec:asymmetrical}

In this model we fix the eccentricity of the disk $e$ and argument of periastron $\omega_D$ to be 0. In the previous variable disk model we found the phase shift $\phi$ to be consistent with 0 with an uncertainty in phase shift of just $0.002$, we therefore also fix this parameter to 0 for computational purposes. In this model we allow the disk edge to have two different temperatures. We model this in {\tt XRbinary} by using a ``spot" that is allowed to cover an arbitrary width of the edge of the disk, effectively creating a hot and a cool zone on the outer edge of the disk. Physically, this could be produced by the impact of the gas stream on the disk, which causes the region near the impact hot spot to be hotter than the region on the opposite side of the disk. Changes in the mass transfer rate from the secondary can affect the temperature of this ``spot" (e.g. SDSS J123813.73-033933.0; \citealt{Zharikov06}), which could be responsible for the observed year to year variations and the $\sim 796$ day periodicity derived in \S\ref{sec:period}.

In this model we fit for the temperature ratio between the hot and cool side of the disk edge $T_h$, the width of the hot side of the disk edge $W_h$, and the location of the center of the hot edge of the disk $\theta_h$; these last two measured in degrees. $\theta_h$ is defined such that $\theta_h = 0$ deg is the direction pointing from the primary straight away from the secondary, and $\theta_h = 90$ deg points towards the observer at phase 0.75, when the observer sees the side of the disk where we would expect an accretion hot spot to be.

We fit the model in the same way as described in Section~\ref{sec:eccentric}, in this case running the MCMC with 2000 steps and 400 walkers, and also discarding the first 50\% as burn-in. The resulting model has an $\hat{R} < 1.4$. The most likely values are shown in Table \ref{tab:parameters}. We see a correlation between $W_h$ and $T_h$, since a large hot zone can produce a similar light curve to a smaller zone with a higher temperature. These parameters are also correlated with the disk height $H_D$, which together with $W_h$ define the effective area of the hot zone. The best fit disk radius is $\sim 1.5 R_c$ throughout all epochs, and a hot region that covers $\gtrsim 100$ deg of the edge of the disk. Models predict that for the best fit parameters of CX137, a typical hot spot would cover $\lesssim 5$ deg of the edge of the disk \citep{Livio93}. From observations, \citep{Warner95} find spots that cover the range of $14-40$ deg, much smaller than what we measure for CX137.

\begin{longrotatetable}
\begin{deluxetable*}{cccccccccc}
    \tablecaption{Best-fit model parameters for each epoch \label{tab:parameters}}
    \tablehead{\colhead{Parameter} & \colhead{Prior}  & \colhead{2010}    & \colhead{2011}       & \colhead{2012}      & \colhead{2013}     &  \colhead{2014}   & \colhead{2015}       & \colhead{2016a}     & \colhead{2016b}}
    \startdata
    \rule{0pt}{4ex} \\
    \multicolumn{3}{l}{Model 1 : Variable Disk} \\
    \hline
    $\phi^\dagger$                 & $[-0.1 - 0.1] $  & $0.001\pm 0.002$  &   \nodata            &    \nodata          &    \nodata         & \nodata           &   \nodata            &   \nodata           & \nodata         \\
    $T_E [K]^\dagger$              & $[500 - 5000] $  & $2301^{+2000}_{-500}$ & \nodata          &    \nodata          &    \nodata         & \nodata           &   \nodata            &   \nodata           & \nodata         \\
    $\log({\rm L_D / [erg / s]})$  & $[30 - 35]    $  & $33.44\pm0.05$    &    $33.68\pm0.11$    &    $33.67\pm0.14$   &    $34.06\pm0.3$   & $33.52\pm0.08$    &    $33.46\pm0.04$    &    $33.46\pm0.46$   & $33.57\pm0.14$  \\
    $e_D$                          & $[0.0 - 0.9]  $  & $0.15\pm0.04$     &    $0.49\pm0.03$     &    $0.52\pm0.03$    &    $0.58\pm0.05$   & $0.11\pm0.07$     &    $0.21\pm0.03$     &    $0.46\pm0.11$    & $0.08\pm0.05$   \\
    $\omega_D (\rm deg)$           & $[0 - 360]    $  & $12.21\pm3.58$    &    $108.13\pm2.9$    &    $96.21\pm4.49$   &    $83.15\pm8.74$  & $17.85\pm19.97$   &    $100.63\pm7.09$   &    $73.14\pm4.67$   & $78.12\pm21.11$ \\
    $R_D [R_c]$                    & $[1.0 - 3.0]  $  & $1.30\pm0.03$     &    $1.03\pm0.03$     &    $1.01\pm0.02$    &    $1.02\pm0.03$   & $1.48\pm0.10$     &    $1.99\pm0.09$     &    $1.33\pm0.23$    & $2.10\pm0.21$   \\
    $H_D [a]$                      & $[0.005 - 0.1]$  & $0.009\pm0.002$   &    $0.029\pm0.001$   &    $0.028\pm0.001$  &    $0.027\pm0.001$ & $0.019\pm0.003$   &    $0.046\pm0.003$   &    $0.041\pm0.005$  & $0.041\pm0.004$ \\
    $f^*$                          &                  & $0.52\pm0.02$     &    $0.48\pm0.03$     &    $0.52\pm0.03$    &    $0.49\pm0.03$   & $0.49\pm0.02$     &    $0.52\pm0.02$     &    $0.55\pm0.03$    & $0.48\pm0.02$   \\
    \rule{0pt}{4ex} \\
    \multicolumn{3}{l}{Model 2 : Asymmetrical Disk Brightness} \\
    \hline
    $\log({\rm L_D / [erg / s]})$  & $[30 - 35]    $  & $32.94\pm0.09$    &    $33.06\pm0.10$    &    $33.19\pm0.20$   &    $32.99\pm0.2$   & $33.16\pm0.08$    &    $32.95\pm0.08$    &    $32.90\pm0.25$   & $33.11\pm0.16$  \\
    $R_D [R_c]$                    & $[1.0 - 3.0]  $  & $1.45\pm0.16$     &    $1.57\pm0.08$     &    $1.52\pm0.17$    &    $1.52\pm0.16$   & $1.43\pm0.10$     &    $1.56\pm0.14$     &    $1.47\pm0.18$    & $1.36\pm0.13$   \\
    $H_D [a]$                      & $[0.005 - 0.1]$  & $0.027\pm0.002$   &    $0.028\pm0.003$   &    $0.030\pm0.003$  &    $0.028\pm0.004$ & $0.026\pm0.007$   &    $0.028\pm0.006$   &    $0.029\pm0.006$  & $0.029\pm0.007$ \\
    $T_E$                          & $[500, 5000]  $  & $1664 \pm 180$    &    $1669 \pm 104$    &    $1772 \pm 97$    &    $1762 \pm 113$  & $1647 \pm 203$    &    $1706 \pm 182$    &    $1671 \pm 150$   & $1474 \pm 132$  \\
    $T_h$                          & $[1.0 - 10.0] $  & $1.97\pm0.20$     &    $2.23\pm0.16$     &    $2.27\pm0.13$    &    $2.33\pm0.17$   & $2.08\pm0.24$     &    $2.03\pm0.20$     &    $2.13\pm0.18$    & $2.04\pm0.10$   \\
    $\theta_h [deg]$               & $[0.0 - 180]  $  & $10.7\pm1.7$      &    $102.3\pm0.7$     &    $88.3\pm0.7$     &    $61.7\pm0.5$    & $13.4\pm1.6$      &    $150.0\pm1.1$     &    $70.3\pm1.5$     & $100.9\pm3.1$   \\
    $W_h [deg]$                    & $[0.0 - 300 ] $  & $149.7\pm3.4$     &    $177.2\pm10.6$    &    $98.4\pm2.2$     &    $168.3\pm0.004$ & $249.9\pm16.8$    &    $188.3\pm3.8$     &    $196.9\pm5.1$    & $119.0\pm19.0$  \\
    $f^*$                          &                  & $0.58\pm0.02$     &    $0.53\pm0.02$     &    $0.49\pm0.02$    &    $0.55\pm0.02$   & $0.49\pm0.02$     &    $0.58\pm0.02$     &    $0.60\pm0.02$    & $0.52\pm0.02$   \\
    \rule{0pt}{4ex} \\
    \multicolumn{3}{l}{Model 3 : Spotted Secondary} \\
    \hline
    $R_D [R_c^\dagger]$            & $[1.0 - 3.0]  $  & $1.48\pm 0.01$    &    \nodata           &    \nodata          &    \nodata         & \nodata           &   \nodata            &   \nodata           & \nodata         \\
    $H_D [a]^\dagger$              & $[0.005 - 0.1]$  & $0.030\pm0.001$   &    \nodata           &    \nodata          &    \nodata         & \nodata           &   \nodata            &   \nodata           & \nodata         \\
    $\log({\rm L_D / [erg / s]})$  & $[30 - 35]    $  & $33.65\pm0.07$    &    $32.74\pm0.01$    &    $33.57\pm0.09$   &    $33.49\pm0.08$  & $33.70\pm0.06$    &    $33.22\pm0.07$    &    $33.43\pm0.09$   & $33.47\pm0.07$  \\
    $T_E$                          & $[500 - 5000] $  & $2223 \pm 150$    &    $4509 \pm 21 $    &    $2959 \pm 122$   &    $3007 \pm 113$  & $2358 \pm 185$    &    $2363 \pm 137$    &    $2239 \pm 201$   & $2392 \pm 165$  \\
    $\theta_{S1}$                  & $[0 - 250]    $  & $170.4 \pm 2.5$   &    $68.2 \pm 1.6$    &    $153.1 \pm 3.1$  &    $162.9 \pm 2.8$ & $163.1 \pm 5.4$   &    $51.7 \pm 1.9$    &    $146.4 \pm 2.2$  & $143.1 \pm 7.2$ \\
    $\phi_{S1}$                    & $[-110 - 110] $  & $-80.8 \pm 3.1$   &    $-55.2 \pm 2.6$   &    $-99.9 \pm 2.4$  &    $-91.1 \pm 4.3$ & $-94.0 \pm 42.9$  &    $-45.5 \pm 3.5$   &    $-93.6 \pm 1.9$  & $-88.3 \pm 5.9$ \\
    $R_{S1}$                       & $[0.0 - 20.0] $  & $12.5 \pm 0.4$    &    $13.7 \pm 1.4$    &    $16.0 \pm 0.9$   &    $16.9 \pm 0.7$  & $12.7 \pm 1.5$    &    $14.3 \pm 0.9$    &    $16.3 \pm 1.1$   & $9.5 \pm 2.8$   \\
    $T_{S1}$                       & $[0.1 - 1.0]  $  & $0.5 \pm 0.2$     &    $0.4 \pm 0.2$     &    $0.4 \pm 0.1$    &    $0.4 \pm 0.2$   & $0.5 \pm 0.2$     &    $0.5 \pm 0.1$     &    $0.3 \pm 0.2$    & $0.6 \pm 0.3$   \\
    $\theta_{S2}$                  & $[0 - 250]    $  & $64.2 \pm 2.9$    &    $88.2 \pm 0.3$    &    $80.9 \pm 1.6$   &    $88.5 \pm 1.5$  & $109.9 \pm 8.4$   &    $62.1 \pm 1.2$    &    $94.2 \pm 1.3$   & $99.3 \pm 2.8$  \\
    $\phi_{S2}$                    & $[70 - 290 ]  $  & $209.4 \pm 5.3$   &    $159.9 \pm 1.7$   &    $116.8 \pm 2.4$  &    $157.4 \pm 3.9$ & $159.4 \pm 18.6$  &    $174.5 \pm 4.7$   &    $157.4 \pm 3.8$  & $160.9 \pm 8.7$ \\
    $R_{S2}$                       & $[0.0 - 20.0] $  & $6.9 \pm 0.7$     &    $12.9 \pm 0.2$    &    $16.1 \pm 0.7$   &    $14.4 \pm 0.2$  & $4.1 \pm 0.5$     &    $10.7 \pm 0.5$    &    $10.9 \pm 0.2$   & $6.8 \pm 0.4$   \\
    $T_{S2}$                       & $[0.1 - 1.0]  $  & $0.5 \pm 0.2$     &    $0.5 \pm 0.1$     &    $0.4 \pm 0.1$    &    $0.5 \pm 0.1$   & $0.4 \pm 0.2$     &    $0.6 \pm 0.2$     &    $0.4 \pm 0.2$    & $0.6 \pm 0.2$   \\
    $f^*$                          &                  & $0.43\pm0.02$     &    $0.69\pm0.02$     &    $0.46\pm0.02$    &    $0.49\pm0.02$   & $0.41\pm0.02$     &    $0.59\pm0.02$     &    $0.51\pm0.02$    & $0.50\pm0.02$   \\
    \enddata
    \tablecomments{Best model parameters and 1$\sigma$ error bars for the realizations shown in Figure~\ref{fig:model}. The parameters of the disk are: The orbital phase $\phi$, the luminosity L$_D$, the eccentricity $e_D$, the argument of periastron $\omega_D$, the disk radius $R_D$ in units of the circularization radius $R_c$, and edge height $H_D$ in units of semi-major axis $a$, and the temperature of the edge of the disk $T_E$. We also include the fractional contribution of the donor star to the total flux of the system $f$, calculated in $V$-band from the posterior distribution of the other parameters from the model. The uncertainties are purely statistical error bars obtained from the posterior distribution of the MCMC. For most parameters we adopt a flat prior, except for the disk luminosity, which is flat in $\log({\rm L_D})$. The disk radius $R_D$ is limited to be less than 0.9 times the Roche Lobe radius of the primary $R_1$.}
    \tablenotetext{*}{These parameters were not fit for, but were calculated using all the posterior distribution samples of the fitted parameters.}
    \tablenotetext{\dagger}{These parameters are kept constant throughout all epochs.}
\end{deluxetable*}
\end{longrotatetable}

We find a best fit for the secondary temperature of $3814\pm20$ K and a secondary mass of $M_2 = 0.68\pm0.03$ M$_\odot$. A cooler but more massive star is not necessarily consistent with the K7 secondary we expect from our spectral analysis in Section~\ref{sec:spectra}. Allowing the disk to be hotter effectively lowered the temperature of the secondary to the point where this model is not entirely self-consistent, and therefore disfavored. This model can help towards a better understanding of the systematic uncertainties in measuring $M_1$, $M_2$, and $i$. Finally, we explore a third model in which the accretion disk is circular and the disk edge has one uniform temperature, but we include two spots on the surface of the secondary.

\subsection{Model 3 : Spotted Secondary}\label{sec:spots}

Finally, we fit the light curves with a model in which the accretion disk is forced to be circular, and have an edge with a single temperature, fixing $W_h = 0$, $\theta_h = 0$, and $T_h = 1$. We place two spots on the surface of the secondary with polar coordinates $\phi_{S1}$, $\theta_{S1}$, $\phi_{S2}$, and $\theta_{S2}$, respectively; and fix $ -110 {\rm\ deg} < \phi_{S1} < 110 {\rm\ deg}$, and $ 70 {\rm\ deg} < \phi_{S2} < 290 {\rm\ deg}$. This prior effectively constrains spot 1 to be on the side of the secondary facing the observer during orbital phase 0.75, and spot 2 on the opposite side of the secondary, allowing for a small overlap region of $20$ deg. The spots have respective angular sizes $R_{S1}$, and $R_{S2}$, and a temperature ratio with respect to the secondary $T_{S1}$, and $T_{S2}$, which are constrained to be $< 1$. We fit for the size and height of the disk as in the previous models, but for computational purposes we constrain them to be the same throughout all epochs. We find that the spotted secondary model requires two spots to be able to explain the fact that the brighter peak at phase 0.75 exhibits larger brightness variations than the dimmer peak at phase 0.25 (See the top panel of Figure~\ref{fig:phased}).

We fit the model in the same way as the one described in Section~\ref{sec:eccentric}, running the MCMC with 2500 steps and 400 walkers, discarding the first 50\% as burn-in. The resulting model has an $\hat{R} < 1.5$. The most likely values are shown in Table \ref{tab:parameters}. We caution that the parameters of the spots are very highly correlated, a small cold spot can produce the same light curve as a large but hotter spot. Nevertheless, the relevant physical parameters such as the mass ratio and inclination appear Gaussian and mostly unaffected by the spot parameters.

We find that $\sim 3$\% of the surface of the secondary is covered by the two modeled spots. For reference, \citep{Watson06} find through Roche Lobe tomography that for the 9.9 hr orbital period CV AE Aqr $\sim 18$\% of one hemisphere of the secondary is spotted. Similarly, the 15 hr orbital period CV BV Cen has $\sim 25$\% of a hemisphere covered by spots \citep{Watson07}.

We determine a best fit secondary temperature of $4050\pm30$ K and a secondary mass of $M_2 = 0.65\pm0.05$ M$_\odot$; very similar to the parameters obtained from Model 1 described in \S\ref{sec:eccentric}. We show the light curve of each epoch, the corresponding most likely model, and the residuals in Figure~\ref{fig:model}. We only include a plot of the spotted secondary model, since all three models presented here are able to reproduce the light curve shape, and visually speaking are effectively indistinguishable. The data are shown phase-folded at the photometric ephemeris with $T_0 = 2455260.8204$ and orbital period $P = 10.34488$ h (derived in \S\ref{sec:period}).

\section{Discussion} \label{sec:params}

\begin{deluxetable*}{ccccc}
    \tablecaption{Best fit parameters \label{tab:models}}
    \tablewidth{0pt}
    \tablehead{\colhead{Parameter} & \colhead{Prior} & \colhead{Variable Disk} & \colhead{Asymmetrical Brightness} & \colhead{Spotted Secondary}}
    \startdata
    $i$             & $\cos([0.0, 90])$ & $63.8\pm0.5$ deg          & $62.2\pm0.2$ deg          & $63.1\pm0.4$ deg          \\
    $T_2$           & $[3500, 4100]$    & $4055\pm25$ K             & $3850\pm50$ K             & $4050\pm30$ K             \\
    $q$             & $0.79 \pm 0.06$   & $0.767\pm0.005$           & $0.78\pm0.01$             & $0.779\pm0.006$           \\
    $v \sin(i)^*$   & $101 \pm 3.0$     & $99.5\pm0.2$ km s$^{-1}$  & $100.6\pm0.1$ km s$^{-1}$ & $100.5\pm0.2$ km s$^{-1}$ \\
    ${M_1}^*$       & \nodata           & $0.81\pm0.05$ M$_\odot$   & $0.86\pm0.03$ M$_\odot$   & $0.83\pm0.05$ M$_\odot$   \\
    ${M_2}^*$       & \nodata           & $0.62\pm0.04$ M$_\odot$   & $0.68\pm0.03$ M$_\odot$   & $0.65\pm0.05$ M$_\odot$   \\
    ${R_{\rm 2}}^*$ & \nodata           & $0.92\pm0.09$ R$_\odot$   & $1.02\pm0.07$ R$_\odot$   & $0.97\pm0.10$ R$_\odot$   \\
    \enddata
    \tablecomments{List of the best fit parameters that are constant throughout all epochs of photometry and fit for in all models. $i$ is the orbital inclination, $T_2$ is the secondary temperature, $q$ is the mass ratio, $v \sin(i)$ is the secondary's rotational velocity, and $M_1$ and $M_2$ are the primary and secondary mass, respectively. And ${R_{\rm 2}}^*$ is the radius of the Roche Lobe of the secondary. For most fit for parameters we adopt a flat prior, except for the orbital inclination, which is flat in $\cos(i)$, and the mass ratio, which has a Gaussian prior.}
    \tablenotetext{*}{These parameters were not fit for, but were calculated using all the posterior distribution samples of the fitted parameters.}
\end{deluxetable*}

\subsection{Stellar Parameters}

We calculate $f$ for each model by measuring the relative flux fraction that the secondary contributes to the total flux of the system in the $V$-band, the closest band to the $ 5580 - 6150$ \AA\ wavelength range used in Section~\ref{sec:spectra} to derive $f = 0.52 \pm 0.06$ from the spectroscopy. From the light curve modeling we find $f$-factors averaged over a full orbit for all epochs of photometry of: $f = 0.50 \pm 0.03$ for the variable disk model, $f = 0.54 \pm 0.04$ for the asymmetrical edge brightness model and $f = 0.51 \pm 0.09$ for the spotted star model. Most of these are in perfect agreement with the value measured from the spectra. The $f$ as a function of epoch is shown in Table~\ref{tab:parameters}.

We find best-fit values for the primary mass of \mbox{$M_1 = 0.81$ M$_\odot$}, \mbox{$M_1 = 0.86$ M$_\odot$}, and \mbox{$M_1 = 0.83$ M$_\odot$} for models 1, 2, and 3, respectively. The statistical uncertainties reported in Table~\ref{tab:models} are in the order of the systematic uncertainties from assuming different models. Accounting for these, we adopt a primary mass estimate of \mbox{$M_1 = 0.83 \pm 0.06$ M$_\odot$}, typical for white dwarfs in CVs (e.g. $M_{\rm WD} = 0.83 \pm 0.23$ M$_\odot$; \citealt{Zorotovic11}), and too small for a neutron star \citep{Ozel16}. The best estimate for the mass for the secondary is \mbox{$M_2 =0.65\pm0.07$ M$_\odot$}, consistent with the mass of a main sequence K7 star \citep{Cox00} and in agreement with the best fit template match found in Section~\ref{sec:spectra}. We find a best fit radius for the Roche Lobe of the secondary of $R_2 = 0.97 \pm 0.15$ R$_\odot$. This radius is larger than expected for a main sequence K7 star (which have typical values of $R \sim 0.65$ R$_\odot$; \citealt{Pecaut13}), supporting an evolved secondary in CX137.

\begin{figure}
	\centering
	\includegraphics[width=\columnwidth]{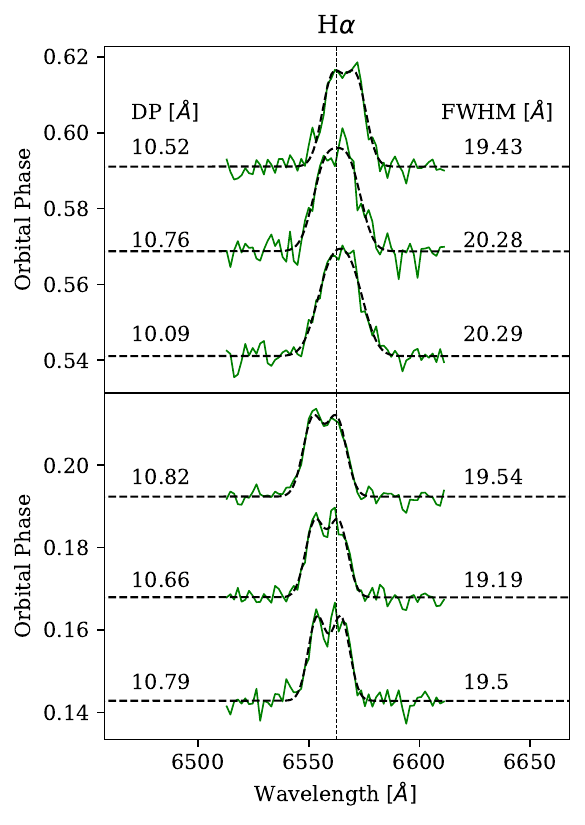}
	\caption{Emission line profiles for H$\alpha$ at 6 different phases. The best-fit separation between the two peaks (DP) and the FWHM is shown in each panel. We determine a ratio of DP$/$FWHM$=0.55 \pm 0.02$ following the methods of \cite{Casares16}. \label{fig:halpha}}
\end{figure}

\begin{figure}
	\includegraphics[clip,width=0.90\columnwidth]{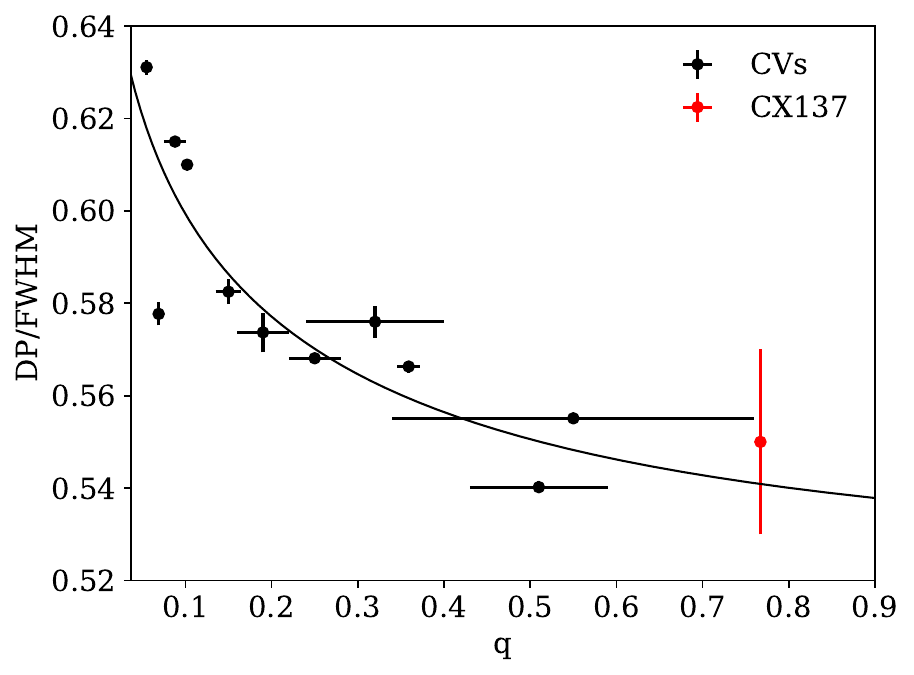}
	\caption{Relation between mass ratio $q$ and ratio of peak separation DP to FWHM of H$\alpha$ for known CVs. The black line is an empirical relation found in \cite{Casares16}, from which this figure is adapted. The parameters found for CX137, shown in red, are consistent with the existing relations for CVs. Error bars not visible are smaller than the marker size. \label{fig:casares}}
\end{figure}

From the spectra, we determine the ratio of the double-peak separation (DP) to the full width half maximum (FWHM) of the H$\alpha$ emission line following the method of \cite{Casares16}. We fit H$\alpha$ with a double Gaussian function to measure the DP between the two line peaks DP$ = 484 \pm 12$ km s$^{-1}$ and then fit a single Gaussian to determine the FWHM$ = 901 \pm 19$ km s$^{-1}$. We find the average ratio to be DP$/$FWHM$ = 0.55 \pm 0.02$, the result of these fits are shown in Figure~\ref{fig:halpha}. In Figure~\ref{fig:casares} we show the $q$, and DP/FWHM of H$\alpha$ plotted alongside the values for other known CVs. We confirm that our parameter estimates agree well with the $q-$DP/FWHM relation for CVs determined by \cite{Casares16}. \cite{Torres14} suggested the double-peaked structure of H$\alpha$ might be due to contamination from photospheric absorption lines from the secondary (e.g., \citealt{Torres19}). Nevertheless, the values we derived for CX137 agree with this trend, and strengthens the case that CX137 is a CV.

Similarly, we calculate the expected FWHM of H$\alpha$ using the FWHM-$K_2$ relation for CVs from \cite{Casares15}. A mass ratio $q = 0.78$ and a FWHM $ = 936 \pm 35 $ km s$^{-1}$, corresponds to an expected value of $K_2$ = $145 \pm 22$ km s$^{-1}$, where the uncertainty is dominated by the scatter from the Casares relation. Consistent within the measured value of $K_2$ = $161 \pm 6$ km s$^{-1}$.

We measure the systemic velocity of CX137 from the optical spectra to be \mbox{$\gamma = 54 \pm 4$ km s$^{-1}$} (Figure~\ref{fig:rvfit}). Given the proper motion and distance to CX137 obtained by \textit{Gaia} (See section~\ref{subsec:gaia}), we can determine the space velocity of CX137 with respect to the Sun to be \mbox{$v = 62 \pm 4$ km s$^{-1}$}, statistically consistent with other CVs (\mbox{$v = 51 \pm 7$ km s$^{-1}$}; \citealt{Ak10}).

In addition to the orbital period of the binary, we detect a tentative periodicity of $\sim 796$ days. Stellar spots are known to live well over this amount of time and up to $\sim 10$ years (e.g, \citealt{Hall94, Savanov14}). As we saw in Model 3, it is possible that these long-term periodicity is produced by the evolution of spots on the surface of the secondary. Nevertheless, our photometry only covers a baseline three times that of this period, making its interpretation or physical origin hard to establish.

\subsection{X-ray Luminosity}

\cite{Torres14} provide a lower limit to the X-ray luminosity of CX137 of \mbox{$L_x > 5.8\times10^{30}$ erg s$^{-1}$} for a distance of 0.7 kpc and assuming a hydrogen column density $N_H = 10^{21}$ cm$^{-2}$. Here, we improve this measurement by using the distance to CX137 from \textit{Gaia} of \mbox{$d = 879^{+59}_{-52}$ pc} \citep{Bailer18}. In addition, we calculate the extinction in the line of sight to CX137 from the Bayestar19 3D dust maps \citep{Green19} to be $A_V \approx 0.58$. We obtain an $N_H = 1.7\times10^{21}$ cm$^{-2}$ from the $A_V$--$N_H$ relation from \cite{Guver09}. We calculate a counts to unabsorbed flux conversion ratio of $5.6\times10^{-15}$ erg cm$^{-2}$ s$^{-1}$ count$^{-1}$ for a 2.16ks exposure taken with ACIS-I during \textit{Chandra} Cycle 9, using a power-law spectrum with $\Gamma = 2$. This corresponds to a $0.5-10$ keV unabsorbed flux of $(8.4 \pm 2.1)\times10^{-14}$ erg cm$^{-2}$ s$^{-1}$, or $L_x = (7.8 \pm 2.2) \times10^{30}$ erg s$^{-1}$ at the distance from \textit{Gaia}.

We can estimate an accretion rate following the method of \cite{Beuermann04} using $L_{\rm acc} = \dot{M} G M_1 (1/R_1 - 1/R_D)$, $R_1 = (1.463 - 0.885 (M_1 / M_\odot)) \times 10^9$ cm, and $L_{\rm acc} = (1 + \alpha) L_x$; where $\alpha$ is typically 0.1. We adopt our best estimate for the primary mass of $M_1 = 0.83$ M$_\odot$, and a typical disk radius of $R_D = 10^{10}$ cm, as determined by our models presented in \S\ref{sec:modeling}. We obtain an accretion rate estimate of $\dot{M} \sim 10^{15}$ g s$^{-1}$ ($10^{-10.8}$ M$_\odot$ yr$^{-1}$).

\cite{Bahramian20} detected CX137 at a higher $L_x$ in the \textit{Swift} Bulge Survey \citep{Shaw20} during repeated biweekly scans of the Galactic Bulge with the Neil Gehrels \textit{Swift} Observatory. They measured an average $L_x=5\times10^{31}$ erg s$^{-1}$ over many epochs in 2017, with a peak of $L_x = 3\pm2\times10^{32}$ erg s$^{-1}$, indicating a flux increase of $38^{+33}_{-26}$ compared to the \textit{Chandra} GBS measurement, which would consequently bring up the accretion rate to $\dot{M} \sim 4\times10^{16}$ g s$^{-1}$ ($10^{-9.2}$ M$_\odot$ yr$^{-1}$) during this period. \cite{Teeseling96} found that the accretion rate in non-magnetic CVs is likely underestimated by a factor of $\sim 2$ for systems with inclinations of $\gtrsim 60$ deg. This would bring the accretion rate to $\dot{M} \sim 10^{17}$ g s$^{-1}$ ($10^{-8.8}$ M$_\odot$ yr$^{-1}$), more similar to the $\dot{M}$ expected for a Roche Lobe filling subgiant with an orbital period of $10$ hr \citep{King96}. An accretion rate of $\dot{M} \sim 10^{17}$ g s$^{-1}$ is expected for CVs with long periods $\gtrsim 4$ hr, yet it is still low for a CV with a 10 hr period like CX137 \citep{Wynn97}.

Using the $L_x$ vs. duty cycle correlation for dwarf novae from \citet{Britt15} we can estimate the duty cycle for CX137 to be $0.063 \pm 0.022$. Accounting for observational cadence, the source should have been in outburst during $94\pm34$ days out of the 1,504 days CX137 was observed by OGLE. One explanation for the lack of outbursts might be that CVs with long orbital periods tend to have shorter outbursts \citep{Hameury17}. Given that the secondary star in CX137 contributes a large fraction of the total flux, an outburst would be of low amplitude, and we could have missed it if it happened when the source was not being observed. KIC 5608384 is another example of a CV with a long period (8.7 h) and a low accretion rate ($\dot{M} = 0.3 - 6.5 \times 10^{-9}$ M$_\odot$ yr$^{-1}$) that showed only one 4 day outburst in four years of Kepler photometry \citep{Yu19}.

\section{Conclusion} \label{sec:conclusions}

We obtained multiple spectra of the binary star CX137 and constructed a radial velocity curve from which we measure a $K_2 = 161.1 \pm 0.7$ km s$^{-1}$ and a systemic velocity \mbox{$\gamma = 54 \pm 4$ km s$^{-1}$}. Additionally, we modeled 7 years of optical photometry. The optical light curve has an asymmetrical sine curve shape, which we interpret as being due to ellipsoidal modulations of a tidally distorted secondary star. We see long-term variations in the shape of the light curve, which are well fit by a spotted secondary star (Model 3; \S\ref{sec:spots}). From the light curve modeling we derive a best fit inclination of \mbox{$i = 63.0\pm0.7$ deg}, a primary mass of \mbox{$M_1 = 0.83 \pm 0.06$ M$_\odot$}, consistent with a white dwarf accretor, and a secondary mass of \mbox{$M_2 = 0.65 \pm 0.07$ M$_\odot$}, consistent with an evolved K7 secondary.

\acknowledgments
This project was supported in part by an NSF GROW fellowship. PGJ and ZKR acknowledge funding from the European Research Council under ERC Consolidator Grant agreement no 647208. JS was supported by a Packard Fellowship. MAPT acknowledge support by the Spanish MINECO under grant AYA2017-83216-P and support via Ram\'on y Cajal Fellowship RYC-2015-17854. We thank Tom Marsh for the use of {\tt molly}. This work has made use of data from the European Space Agency (ESA) mission {\it Gaia} (\url{https://www.cosmos.esa.int/gaia}), processed by the {\it Gaia} Data Processing and Analysis Consortium (DPAC, \url{https://www.cosmos.esa.int/web/gaia/dpac/consortium}). Funding for the DPAC has been provided by national institutions, in particular the institutions participating in the {\it Gaia} Multilateral Agreement. The ISIS spectroscopy was obtained with the WHT, operated on the island of La Palma by the Isaac Newton Group of Telescopes in the Spanish Observatorio del Roque de los Muchachos of the Instituto de Astrofisica de Canarias. This paper includes data gathered with the 6.5 meter Magellan Telescopes located at Las Campanas Observatory, Chile. This research has made use of NASA’s Astrophysics Data System. This research has made use of the SIMBAD database, operated at CDS, Strasbourg, France. The OGLE project has received funding from the National Science Centre, Poland, grant MAESTRO 2014/14/A/ST9/00121 to AU.

\software{Astropy\citep{astropy18}, PyRAF\citep{science12}, SAOImage DS9 \citep{Smithsonian00}, emcee\citep{Foreman13}, corner \citep{Foreman16}, Matplotlib\citep{hunter07}, SciPy\citep{Walt11}, NumPy\citep{Oliphant07}, extinction(\citep{Barbary16}), PYPHOT(\url{https://github.com/mfouesneau/pyphot}), Molly({\url{http://deneb.astro.warwick.ac.uk/phsaap/software/molly/html/INDEX.html}}), XRbinary(\url{http://www.as.utexas.edu/~elr/Robinson/XRbinary.pdf}).}

\section{Data Availability} \label{sec:data}

All the optical photometry used for this work are available on the online supplementary material version of this article. And the radial velocity data is shown in Table~\ref{tab:radial_velocities}.

\bibliography{references}

\appendix
\setcounter{figure}{0}
\setcounter{table}{0}

\section{Radial Velocity Table}\label{sec:apendix}
This section contains a data table with all the relevant radial velocity measurements.

\begin{deluxetable}{cccc}
	\tablecaption{Radial Velocity Measurements \label{tab:radial_velocities}}
	\tablehead{\colhead{HJD} & \colhead{Phase} & \colhead{Blue Arm}  & \colhead{Red Arm} \\
                    &	                  & \colhead{km s$^{-1}$} & \colhead{km s$^{-1}$}} 
	\startdata
	2457928.59333808 &	$0.12 \pm 0.01  $ & $ 147.58 \pm 17.02$ & $ 176.82 \pm 32.42$ \\
	2457928.60054154 &	$0.14 \pm 0.01  $ & $ 165.92 \pm 17.87$ & $ 174.49 \pm 43.66$ \\
	2457928.61416582 &	$0.17 \pm 0.01  $ & $ 188.48 \pm 15.63$ & $  204.0 \pm 41.87$ \\
	2457928.62136939 &	$0.19 \pm 0.01  $ & $ 191.22 \pm 15.28$ & $ 229.08 \pm 46.91$ \\
	2457928.63017967 &	$0.21 \pm 0.007 $ & $ 279.04 \pm 35.36$ & $ 173.87 \pm 23.30$ \\
	2457928.63866717 &	$0.23 \pm 0.007 $ & $ 269.75 \pm 34.11$ &  \nodata \\
	2457945.52058442 &	$0.26 \pm 0.007 $ & $ 179.98 \pm 34.33$ & $ 143.12 \pm 22.97$ \\
	2457945.52778765 &	$0.28 \pm 0.007 $ & $ 138.03 \pm  7.03$ & $ 170.52 \pm 10.81$ \\
	2457945.53499120 &	$0.30 \pm 0.007 $ & $ 325.63 \pm 35.34$ & $ 173.73 \pm  8.88$ \\
	2457945.54556637 &	$0.32 \pm 0.007 $ & $ 296.82 \pm 25.86$ & $ 176.81 \pm 14.35$ \\
	2457945.55276959 &	$0.48 \pm 0.007 $ & $  51.32 \pm 18.00$ & $   88.6 \pm 20.27$ \\
	2457945.55997296 &	$0.50 \pm 0.007 $ & $  41.01 \pm 19.48$ & $  36.56 \pm 22.01$ \\
	2457955.53460348 &	$0.51 \pm 0.007 $ & $ -10.06 \pm 20.93$ & $  77.07 \pm 22.96$ \\
	2457955.54538190 &	$0.54 \pm 0.007 $ & $ -51.22 \pm 19.54$ & $   61.0 \pm 21.58$ \\
	2457955.55613857 &	$0.56 \pm 0.01  $ & $ -32.42 \pm 19.73$ & $  77.23 \pm 19.12$ \\
	2457955.56690152 &	$0.57 \pm 0.007 $ & $ -65.28 \pm 21.94$ & $  -9.73 \pm 17.04$ \\
	2457955.57773804 &	$0.71 \pm 0.007 $ & $-115.45 \pm 17.05$ & $ -95.98 \pm 11.91$ \\
	2457956.39406596 &	$0.74 \pm 0.01  $ & $-103.73 \pm 16.63$ & $-107.96 \pm 12.11$ \\
	2457956.40482062 &	$0.76 \pm 0.01  $ & $-102.91 \pm 17.06$ & $-100.52 \pm 11.31$ \\
	2457956.41896960 &	$0.79 \pm 0.01  $ & $-107.86 \pm 16.42$ & $ -93.96 \pm 10.07$ \\
	2457956.42972824 &	$0.81 \pm 0.01  $ & $ -90.24 \pm 17.58$ & $ -83.48 \pm 12.03$ \\
	2457956.44047118 &	$0.71 \pm 0.01  $ & $-118.98 \pm 15.49$ & $ -91.92 \pm 12.57$ \\
	2457956.45125232 &	$0.73 \pm 0.01  $ & $-129.45 \pm 16.73$ & $ -95.19 \pm 11.78$ \\
	2457956.46203540 &	$0.77 \pm 0.01  $ & $-107.13 \pm 16.41$ & $-105.07 \pm 10.57$ \\
	2457956.47942908 &	$0.79 \pm 0.01  $ & $-114.13 \pm 16.48$ & $-115.54 \pm 12.78$ \\
	2457956.49018300 &	$0.82 \pm 0.01  $ & $ -82.39 \pm 14.79$ & $ -61.39 \pm 15.10$ \\
	2457956.50090890 &	$0.84 \pm 0.01  $ & $ -83.39 \pm 14.81$ & $ -73.31 \pm 11.55$ \\
	2457993.39230059 &	$0.87 \pm 0.01  $ & $ -67.96 \pm 15.32$ & $ -85.09 \pm 11.04$ \\
	2457993.40297597 &	$0.91 \pm 0.01  $ & $ -25.48 \pm 15.72$ & $ -35.57 \pm 10.69$ \\
	2457993.41365122 &	$0.93 \pm 0.01  $ & $ -29.25 \pm 15.76$ & $  18.58 \pm 12.15$ \\
	2457995.36471471 &	$0.96 \pm 0.01  $ & $   3.35 \pm 18.52$ & $  37.29 \pm 12.94$ \\
	2457995.37539082 &	$0.54 \pm 0.01  $ & $  -5.51 \pm 19.23$ & $ -10.56 \pm 42.03$ \\
	2457995.38606633 &	$0.57 \pm 0.01  $ & $ -30.79 \pm 20.13$ & $  50.13 \pm 56.59$ \\
	2457995.39677799 &	$0.59 \pm 0.01  $ & $ -44.82 \pm 18.25$ & $ -46.38 \pm 41.77$ \\
	\enddata
	\tablecomments{Radial velocity measurements shown in Figure~\ref{fig:rvfit} taken simultaneously with the red and blue arm of the ISIS spectrograph. Corrected for heliocentric velocity.}
\end{deluxetable}

\end{document}